\documentclass[prl,aps,10pt,twocolumn,showpacs,preprintnumbers,superscriptaddress]{revtex4}

\usepackage{amsmath,amssymb,amsfonts,amsthm}
\usepackage{amsbsy} 
\usepackage{epsfig}
\usepackage{latexsym}
\usepackage{dbnsymb}
\input amssym.def
\input amssym.tex

\usepackage{color}
\usepackage{hyperref}

\def\com{\color{magenta}}

\newcommand{\oarX}[1]{\href{http://arxiv.org/abs/#1}{{\ttfamily\com #1}}}
\newcommand{\arX}[1]{\href{http://arxiv.org/abs/#1}{{\ttfamily\com arXiv:#1}}}

\def\barr{\begin{array}}
\def\earr{\end{array}}
\def\half{\frac{1}{2}}
\def\ben{\begin{equation}}
\def\een{\end{equation}}
\def\bs{\begin{subequations}}
\def\es{\end{subequations}}
\def\bena{\begin{eqnarray}}
\def\eena{\end{eqnarray}}

\def\bR{\mathbb{R}}

\def\SO{{\rm SO}}

\def\SU{{\rm SU}}

\def\im{{\rm i}}

\def\M{\mathcal{M}}

\newcommand{\ie}{\textit{i.e.}~}
\newcommand{\eg}{\textit{e.g.}~}

\begin{document}

\title{Cosmology from Group Field Theory Formalism for Quantum Gravity}
\author{Steffen Gielen}
\email{sgielen@perimeterinstitute.ca}
\affiliation{Riemann Center for Geometry and Physics, Leibniz Universit\"at Hannover, Appelstra\ss e 2, 30167 Hannover, Germany, EU}
\affiliation{Perimeter Institute for Theoretical Physics, 31 Caroline Street North, Waterloo, Ontario N2L 2Y5, Canada} 
\author{Daniele Oriti}
\email{doriti@aei.mpg.de}
\affiliation{Max Planck Institute for Gravitational Physics (Albert Einstein Institute), Am M\"uhlenberg 1, 14476 Golm, Germany, EU}
\author{Lorenzo Sindoni}
\email{sindoni@aei.mpg.de}
\affiliation{Max Planck Institute for Gravitational Physics (Albert Einstein Institute), Am M\"uhlenberg 1, 14476 Golm, Germany, EU}

\begin{abstract}

We identify a class of condensate states in the group field theory (GFT) formulation of quantum gravity that can be interpreted as macroscopic homogeneous spatial geometries. We then extract the dynamics of such condensate states directly from the fundamental quantum GFT dynamics, following the procedure used in ordinary quantum fluids. The effective dynamics is a nonlinear and nonlocal extension of quantum cosmology. We also show that any GFT model with a kinetic term of Laplacian type gives rise, in a semiclassical (WKB) approximation and in the isotropic case, to a modified Friedmann equation. This is the first concrete, general procedure for extracting an effective cosmological dynamics directly from a fundamental theory of quantum geometry.
\end{abstract}

\date{\today}

\preprint{AEI-2013-051, pi-qg-320}

\maketitle
The main challenge faced by all quantum gravity approaches is to bridge the gap between Planck-scale physics and effective physics at macroscopic scales, to provide testable predictions. In background independent approaches, this is a difficult task because the most natural notion of a vacuum state is one that describes no spacetime at all, while macroscopic geometries should be thought of as 
states with a very large number of quantum geometric excitations \cite{weave, coherent}. 
Next, one needs to extract 
an effective dynamics for such highly excited states and relate it to the usual relativistic dynamics. 
This is an ever harder task, 
and no approach has fully succeeded, despite many interesting results 
\cite{LQG,DT}.  

In this Letter, we put forward a concrete proposal to bridge this gap and, for the first time, extract an effective cosmological dynamics directly from a fundamental 
quantum gravity framework (as opposed to  
a minisuperspace reduction). We use the group field theory (GFT) formalism for quantum gravity \cite{GFTreview}, which is strictly related to loop quantum gravity and spin foam models \cite{LQG}, tensor models \cite{TensorReview}, and dynamical triangulations \cite{DT}. Therefore, the relevance of our results extends well beyond the GFT approach.

After identifying a criterion for discrete geometries 
to approximate continuum ones and 
to be compatible with spatial homogeneity, we propose a class of  
GFT states describing continuum macroscopic homogeneous (but anisotropic) geometries: GFT {\it condensates}, superpositions of $N$-particle states satisfying the criterion for spatial homogeneity at each $N$, which are thus spatially homogeneous to arbitrary accuracy. 
The appearance of macroscopic geometries is captured by a process similar to Bose-Einstein condensation (BEC) of appropriate fundamental quanta, thus realizing the idea of {\it spacetime as a condensate} often advocated in the past \cite{hu,GFTfluid}.

Next, we extract the dynamics of such condensate states (in two interesting cases) directly from the fundamental quantum GFT dynamics, following the procedure used in ordinary quantum fluids. This effective dynamics is shown to have the form of a nonlinear and nonlocal extension of quantum cosmology, similar to the one suggested in Ref. \cite{nonlincosm}. As an example, we show that any GFT model involving a Laplacian kinetic term, as suggested by recent work on GFT renormalization, gives rise in a WKB approximation to an equation describing the classical dynamics of a homogeneous universe, and in the isotropic case to a modified Friedmann equation with corrections
 determined by the fundamental GFT dynamics.  

Our procedure applies to any GFT model incorporating appropriate pregeometric data, such as Refs. \cite{EPRL,fk,BO-Immirzi}, and is thus very general.
It opens a new avenue to getting effective equations for an emergent spacetime geometry from a pregeometric scenario and lends weight to claims that such quantum gravity models correspond to general relativity in a semiclassical continuum approximation. Full details of our calculations and results will be presented in a separate publication.

{\em Group field theory.} --- Group field theories
are quantum field theories on group manifolds (or their Lie algebras), 
with a nontrivial combinatorial structure of quantum states and histories. Their quantum states are in fact four-valent graphs labeled by group or Lie algebra elements, 
 which can be equivalently represented as 3D cellular complexes. The quantum dynamics, in a perturbative expansion around the (``no-space'') vacuum, gives a sum of Feynman diagrams dual to 4D cellular complexes of arbitrary topology. The Feynman amplitudes 
for these discrete histories can be written either as spin foam models \cite{SF} or as simplicial gravity path integrals \cite{DT}. The relation with other approaches to quantum gravity is apparent.

For technical simplicity only, we focus here on the Riemannian case. 
The counterpart of our construction for Lorentzian GFT models is straightforward. 

In this setting, GFTs can be defined in terms of a (complex) bosonic field $\varphi(g_1,g_2,g_3,g_4)$ on $\SO(4)^4$, which can be expanded in annihilation operators:
$
\hat\varphi(g_I)=\sum_\nu \varphi_{\nu}(g_I)\hat{a}_{\nu};
$
using the basic operators $\hat{a}_{\nu}^{\dagger}$, one can then construct the GFT Fock space out of the no-space vacuum $|0\rangle$. A quantum of the GFT field, created by the operator $\hat\varphi^{\dagger}(g_1,\ldots,g_4)$, is interpreted  
as a tetrahedron whose geometry is given by the four parallel transports $g_I$ of the gravitational $\SO(4)$ connection along links dual to its faces.
In this picture, a superposition of $N$-particle states in the GFT corresponds to a spin network with $N$ vertices or a complex with $N$ tetrahedra.
One can use a noncommutative Fourier transform to define the field on conjugate Lie algebra variables
$
\tilde\varphi(B_1,B_2,B_3,B_4) 
$
\cite{BO-Immirzi}. The variables $B_I\in\mathfrak{so}(4)$ are bivectors associated to the faces of the tetrahedron: 
\ben
B^{AB}_{\triangle_{I}} \sim \int_{\triangle_{I}} e^A \wedge e^B\,.
\label{integ}
\een
$e$ is a cotetrad field encoding the simplicial geometry. 

In order to ensure this interpretation, the variables $B_I$ must satisfy two types of conditions. First, {\em simplicity constraints} \cite{SF,fk,EPRL,BO-Immirzi}:
\ben
\exists n^A\in S^3\subset\bR^{4}:\quad\forall\;I \quad n_{A}B^{AB}_{I}=0\,.
\label{simpl}
\een
These impose a restriction on the domain of $\varphi$ to a submanifold of $\SO(4)$, with
different constructions having been proposed \cite{EPRL,fk,BO-Immirzi}.
For example \cite{bcgft}, Eq.~(\ref{simpl}) can be imposed by requiring 
\ben
\varphi(g_1,g_2,g_3,g_4)=\varphi(g_1 h_1,g_2 h_2,g_3 h_3,g_4 h_4)\;\forall\, h_I\in \SO(3),
\label{simpl2}
\een
so that $\varphi$ takes values on four copies of $\SO(4)/\SO(3)\sim S^3\sim\SU(2)$. For GFT models with direct relation to loop quantum gravity \cite{EPRL}, instead, the field dependence is reduced to the diagonal $\SU(2)$ subgroup of $\SO(4)$.

A second condition is invariance under gauge transformations that can be implemented as the invariance
\ben
\varphi(g_1,g_2,g_3,g_4)=\varphi(g_1 h, g_2 h, g_3 h,g_4 h)\quad\forall\; h\in \SO(4).
\label{firstgauge}
\een
In Lie algebra variables, (\ref{firstgauge}) encodes a {\em closure constraint}: the bivectors $B_I$ must close to form a tetrahedron \cite{BO-Immirzi}.

The simplicity constraints imply that there exist vectors $e^A_{i}\in\bR^{4}$ (for $i=1,2,3$) such that for all $i$,
\ben
B_{i}^{AB}={\epsilon_i}^{jk} e_{j}^A e_{k}^B\,.
\label{tetrad}
\een

{\em Approximate geometries and homogeneity.} --- In this second quantized formalism, the $N$-particle state 
\ben
|B_{I(m)}\rangle := \frac{1}{N!}\prod_{m=1}^N \hat{\tilde{\varphi}}^{\dagger}(B_{1(m)},\ldots,B_{4(m)})|0\rangle
\label{state}
\een
is interpreted as a discrete geometry of $N$ tetrahedra with bivectors $B_{I(m)}$ associated to the faces. Assuming closure and simplicity constraints, we parametrize Eq.~(\ref{state}) by $3N$ bivectors $\{B_{i(m)}\}$ $(i=1,\ldots,3,\;m=1,\ldots,N)$ of the form Eq.~(\ref{tetrad}). On this space of bivectors, or alternatively the space of $e^A_{i(m)}$, there is an action of $\SO(4)^N$
\ben
B_{i(m)}\mapsto (h_{(m)})^{-1} B_{i(m)} h_{(m)}\,,\quad e_{i(m)}\mapsto e_{i(m)} h_{(m)}\,.
\label{lortraf}
\een

This corresponds to a local frame rotation. 
The gauge-invariant configuration space for each tetrahedron is six-dimensional and may be parametrized by 
\ben
g_{ij(m)} = e_{i(m)}^A e_{Aj(m)}\,.
\label{metric}
\een
Defining the six bilinears $\tilde{B}_{ij}:=B^{AB}_{i}B_{jAB}$, we can express the components $g_{ij}$ in terms of the bivectors $B_{i(n)}$:
\ben
g_{ij}=\frac{1}{8\,{\rm tr }(B_1 B_2 B_3)}{\epsilon_i}^{kl}{\epsilon_j}^{mn}\tilde{B}_{km}\tilde{B}_{ln}\,.
\label{tetmetric}
\een
To associate to these data an approximate continuum (spatial) geometry,
we think of the tetrahedra as embedded into a three-dimensional topological manifold $\M$ on which a Lie group $G$ acts transitively, so that $\M\simeq G/X$, where $X$ can be a discrete subgroup of $G$; $G$ defines the notion of homogeneity \cite{cosmobook}. An embedding of each tetrahedron is specified by the location of one of the vertices and three tangent vectors specifying the directions of the three edges emanating from this vertex
\ben
\tetrahedron_m\,\mapsto\,\left\{x_m \in \M,\;\left\{{\bf v}_{1(m)},{\bf v}_{2(m)},{\bf v}_{3(m)}\right\}\subset T_{x_m}\M\right\}.
\een
In order to exponentiate the tangent vectors to obtain the location of the other three vertices, we can use the Maurer-Cartan connection on $G$ pulled back to $\M$.

We interpret the $\bR^{4}$ vectors $e^A_{i(m)}$ associated to a tetrahedron as physical tetrad vectors integrated along the edges specified by ${\bf v}_{i(m)}$, a natural choice for which is a basis of left-invariant vector fields on $G$:
$
{\bf v}_{i(m)} = {\bf e}_i(x_m),
$
where $\{{\bf e}_i\}$ are the vector fields on $\M$ obtained by push forward of a basis of left-invariant vector fields on $G$.
This requires assuming that the tetrahedra are  associated to regions in the embedded manifold which are sufficiently flat, so that we can approximate the tetrad as constant.

Within this approximation, the vectors $e^A_{i(m)}$ are related to physical tetrad vectors by
$
e^A_{i(m)}=e^A(x_m)({\bf e}_i(x_m))\,.
$
For the $\SO(4)$ invariant quantities $g_{ij}$, 
\ben
g_{ij(m)}=g(x_m)({\bf e}_i(x_m),{\bf e}_j(x_m))\,;
\label{physmet}
\een
thus $g_{ij(m)}$ are the metric components in the frame $\{{\bf e}_i\}$. 

Using the transitive action of $G$, we can say that a discrete geometry of $N$ tetrahedra, specified by the data $g_{ij(m)}$, is {\em compatible with spatial homogeneity} if
\ben
g_{ij(m)}=g_{ij(k)}\quad\forall\; k,m=1,\ldots,N.
\label{homocri}
\een
This criterion only uses intrinsic geometric data and depends on the embedding information only through the choice of $G$.
The correspondence between $N$-particle GFT states and continuum geometries can be viewed as the result of sampling the metric at $N$ points. 

{\it GFT condensates as continuum homogeneous geometries.} --- The GFT framework now allows us to take two more crucial steps: (1) lift the above construction to the quantum setting, and (2) take $N$ as variable and send it to infinity. 
The quantum counterpart of the classical homogeneity condition becomes the requirement that the GFT $N$-particle state has a product structure in which {\it the same wave function}, 
invariant under the transformation \eqref{lortraf}, is assigned to each GFT quantum. Then, arbitrary superpositions of such $N$-particle states can be considered, with $N$ arbitrarily large. 
Notice that, if Eq.~(\ref{homocri}) holds {\em for any $N$}, the reconstructed spatial geometry is homogeneous to arbitrary accuracy.
This is nothing else than 
a GFT quantum condensate state
\cite{GFTfluid}.

We now give two explicit examples of such GFT condensates.
The simplest is a ``single-particle'' condensate,
\ben\label{simple}
|\sigma\rangle := \exp\left(\hat\sigma\right)|0\rangle \quad\text{with}\quad \hat\sigma := \int d^4 g\; \sigma(g_I)\hat\varphi^{\dagger}(g_I)
\een
if we require $\sigma(kg_I)=\sigma(g_I),\; \forall\, k\in \SO(4)$
in addition to Eq.~(\ref{firstgauge}).
The second uses a two-particle operator \bena
|\xi\rangle &:=& \exp\left(\hat\xi\right)|0\rangle \quad \text{with} \\  \quad \hat\xi &:=& 
\half\int d^4 g\;d^4 h\; \xi(g_I h_I^{-1})\hat\varphi^{\dagger}(g_I)\hat\varphi^{\dagger}(h_I)\, ,
\label{dipolestate}
\eena
where, thanks to (\ref{firstgauge}) and $[\hat\varphi^{\dagger}(g_I),\hat\varphi^{\dagger}(h_I)]=0$, the function $\xi$ \emph{automatically} satisfies $\xi(g_I)=\xi(kg_Ik'), \, \forall k,k' \in \SO(4)$. $\xi$ is a ``dipole'' function \cite{dipoleLQC} with the same geometric data as \eqref{simple}, but with the advantage of naturally having the right gauge invariance 
and of taking into account some limited multiparticle correlation. 

{\em Effective cosmological dynamics.} --- We now extract the effective dynamics for a homogeneous quantum space, \ie for GFT condensates, from a generic GFT for 4D quantum gravity, following closely the standard {procedures} used in quantum fluids \cite{BEC}. The action consists of a quadratic (kinetic) term and a higher order interaction:
\ben
S[\varphi]=\int d^4 g\,d^4 g'\,\bar\varphi(g_I)\mathcal{K}(g_I,g'_I)\varphi(g'_I)+\lambda\mathcal{V}[\varphi,\bar\varphi]
\een
leading to the fundamental quantum equation of motion
\ben
\int d^4 g'\,\mathcal{K}(g_I,g'_I)\hat{\varphi}(g'_I)+\lambda\frac{\delta \hat{\mathcal{V}}}{\delta\hat\varphi^{\dagger}(g_I)}=0
\label{quanteom}
\een
(and its complex conjugate). We now apply these operator equations to GFT condensates.
For the two choices of GFT states, we get an effective equation for the ``collective cosmological wave functions'' $\sigma$ or $\xi$.
The simplest effective dynamics is obtained for the states \eqref{simple}.
Since $|\sigma\rangle$ is an eigenstate of $\hat\varphi(g_I)$, the expectation value of Eq.~(\ref{quanteom}) in $|\sigma\rangle$ is the nonlinear equation for $\sigma$:
\ben
\int d^4 g'\,\mathcal{K}(g_I,g'_I)\sigma(g'_I)+\lambda\frac{\delta \mathcal{V}}{\delta\bar\varphi(g_I)}\Big|_{\varphi=\sigma}=0\,.
\een

This 
is the GFT analog of the Gross-Pitaevskii equation for BECs. It 
is a nonlinear and nonlocal (on the space of geometries) equation for the collective cosmological wave function $\sigma$, similar to the ones in Refs. \cite{nonlincosm,GFC}.

For the state $|\xi\rangle$, effective dynamics can be extracted by inserting it into the quantum equations for the GFT $N$-point functions.
In general, the result is a coupled set of equations involving 
convolutions of 
$\xi$,
again nonlinear and nonlocal. If the interaction $\mathcal{V}$ is of odd order, because all odd correlation functions vanish, the two terms in Eq.~(\ref{quanteom}) give independent constraints on the function $\xi$. 
The kinetic part alone gives the nontrivial condition 
\ben
\int d^4 g'' \,\hat{\mathcal{K}}(g'_I,g''_I)\xi(g_I {g''_I}^{-1}) = 0\,.
\label{xiequa}
\een
Since Eq.~(\ref{xiequa}) is linear, it could be interpreted as a standard quantum cosmological equation of motion for $\xi$, with Hamiltonian constraint given by $\tilde{\mathcal{K}}$. (\ref{xiequa}) implies that a condensation of correlated pairs of GFT quanta, for GFT models with odd interactions, is only possible if the kinetic operator $\hat{\mathcal{K}}$ admits a nontrivial kernel.  

{\it Effective modified Friedmann equation.}---We now show that any model whose kinetic operator is the Laplace-Beltrami operator on $\SU(2)^4$, together with a ``mass term,'' gives a modified Friedmann equation in the semiclassical and isotropic limit. $\SU(2)^4$ is a natural domain for many GFT models for 4D gravity, while the Laplacian seems to be required by GFT renormalization 
\cite{GFTrenorm}.

The effective cosmological dynamics reduces to (\eg in a weak-coupling limit, for $| \sigma\rangle$) or contains (for the state
$|\xi\rangle$, which we use in the following) the equation
\ben
\left(\Delta_{g_I}+\mu\right)\xi(g_I {g'_I}^{-1}) = 0\,.
\label{xiequation}
\een

Using the parametrization for $\SU(2)$ given by
$
g = \sqrt{1-\vec{\pi}^2}\,{\bf 1} - \im\vec{\sigma}\cdot\vec\pi\,,\;|\vec{\pi}|\le 1\,,
$
where $\sigma^i$ are the Pauli matrices, the Laplace-Beltrami operator on $\SU(2)$ is
\ben
\Delta_g f(\pi[g]) = \left(\delta^{\alpha\beta}-\pi^\alpha\pi^\beta\right)\partial_\alpha\partial_\beta f(\pi)\,.
\een
Rewriting $\xi(\pi_I[g_I])=A[\pi_I]\exp(\im S[\pi_I]/\kappa)$ 
and taking the eikonal limit
$\kappa \rightarrow 0$, this equation reduces to
\ben
\sum_I\left[B_I\cdot B_I - (\pi_I\cdot B_I)^2\right]
\approx 0,
\label{friedmann}
\een
where the multiplication dot represents the Killing form on $\mathfrak{su}(2)$ and $B_I:=\partial S/\partial \pi_I$  is the momentum conjugate to $\pi_I$. Since $S[\pi(g_I)]=S[\pi(kg_Ik')], \, \forall k,k'  \in \SU(2)$ the $B_I$ satisfy additional relations. Within this WKB approximation, Eq.~(\ref{friedmann}) becomes the Hamilton-Jacobi equation for the classical action $S$. 

In order to identify the $B_I$ and conjugate $\pi_I$ with cosmological variables, we follow their geometric
interpretation as bivectors and conjugate infinitesimal holonomies, and write $B_I=a_I^2\,T_I$ and $\pi_I = p_I V_I$, where each $T_I$ and $V_I$ is a dimensionless normalized Lie algebra element. 
Then, Eq.~(\ref{friedmann}) becomes
\ben
\sum_I a_I^4\left(p_I^2\, c_I^2 - 1\right) \approx 0\,,
\label{friedmann2}
\een
where $c_I=T_I\cdot V_I$ depend on the state. Specializing to an isotropic geometry, we can set $a_I=\gamma_I a,\,p_I=\beta_I p$ for constants $\gamma_I$ and $\beta_I$, and Eq.~(\ref{friedmann2}) becomes
\ben
p^2 - k = O\left(\frac{\kappa}{a^2}\right)\,,
\label{friedmann3}
\een
where $k=\left(\sum_I \gamma_I^4\right)/\left(\sum_I \gamma_I^4 \beta_I^2 c_I^2\right)$. At leading order, this is the classical Friedmann equation for an empty universe with spatial curvature $k$. Since $k>0$, this interpretation is consistent when $G=\SU(2)$. The 
corrections to (\ref{friedmann3}) include both the subdominant terms in the WKB approximation 
and the corrections coming from the higher order terms in the effective cosmological dynamics.

{\em Discussion.} --- This Letter illustrates a new, concrete avenue for extracting an effective cosmological dynamics from  
fundamental quantum gravity. We believe it is the first time that such a direct path is open in background independent, pregeometric quantum gravity approaches. 

The results presented can be summarized as follows. We have identified quantum GFT states (easily exportable to loop quantum gravity 
or simplicial gravity) that are natural candidates to describe
homogeneous (anisotropic) cosmological geometries. 
Similar states have indeed been proposed in 
related contexts \cite{alesci,edward,dipoleLQC}; but, contrary to those proposals, the GFT condensates  
do not depend on any single lattice structure. 
The advantage of this will appear once moving away from homogeneity: inhomogeneities can be encoded in fluctuations above GFT condensate states, and such coherent states support such perturbations at any approximation scale. 
Most importantly, we could extract an effective cosmological dynamics from the fundamental dynamics, in full generality and rather straightforwardly. It takes the form of a nonlinear and nonlocal extension of (loop) quantum cosmology, a GFT analog of Gross-Pitaevskii hydrodynamics in real BECs. This extraction procedure 
can be applied to any given GFT model, specifically to the interesting models proposed in 
Refs. \cite{EPRL,fk,BO-Immirzi}. We have also shown that for any GFT model having a kinetic term of Laplacian form, a modified Friedmann equation can be obtained in the semiclassical and isotropic limit. 

This new avenue points to several directions, all aimed at extracting interesting physics directly from current candidate GFT models for quantum gravity, which are thus solidly rooted in a complete quantum gravity framework,  
for instance, quantum gravity corrections to Friedmann-Robertson-Walker (FRW) cosmology and to the evolution of anisotropies and fluctuations above the GFT condensate that describe inhomogeneities.

At a more formal level, the ongoing work on GFT renormalization \cite{GFTrenorm,vincentTensor} and phase transitions in GFT and tensor models \cite{TensorReview, critical} can now be better directed toward proving rigorously the dynamical realization of the {\it condensation} leading to the states \eqref{simple} or \eqref{dipolestate}. This will establish a solid mathematical basis for the physical picture of continuum space as a GFT condensate.

Research at Perimeter Institute is supported by the Government of Canada through Industry Canada and by the Province of Ontario through the Ministry of Research \& Innovation. The research leading to these results has received funding from the [European Union] Seventh Framework Programme [FP7-People-2010-IRSES] under Grant Agreement No. 269217. S.G. was supported by the Riemann Center for Geometry and Physics. D.O. acknowledges financial support from the A. von Humboldt Stiftung with a Sofja Kovalevskaja Award.

\end{document}